\documentclass[
reprint,
superscriptaddress,
amsmath,amssymb,
aps,
prl,
]{revtex4-2}

\usepackage{graphicx}
\usepackage{dcolumn}
\usepackage{bm}
\usepackage{subcaption}
\usepackage{qcircuit}
\usepackage{braket}
\usepackage{lipsum}
\usepackage{xcolor}

\begin{document}
\title{Mathieu Control of the Effective Coupling in Superconducting Qubits}

\author{Yi-Han Yu}
\thanks{These two authors contributed equally to this work.}
\affiliation{Beijing National Laboratory for Condensed Matter Physics, Institute of Physics, Chinese Academy of Sciences, Beijing 100190, China}
\affiliation{School of Physical Sciences, University of Chinese Academy of Sciences, Beijing 100049, China}

\author{Xin-Yi Li}
\thanks{These two authors contributed equally to this work.}
\affiliation{School of Physical Sciences, University of Chinese Academy of Sciences, Beijing 100049, China}

\author{Kai Xu}
\email{kaixu@iphy.ac.cn}
\affiliation{Beijing National Laboratory for Condensed Matter Physics, Institute of Physics, Chinese Academy of Sciences, Beijing 100190, China}
\affiliation{School of Physical Sciences, University of Chinese Academy of Sciences, Beijing 100049, China}
\affiliation{Beijing Key Laboratory of Fault-Tolerant Quantum Computing, Beijing Academy of Quantum Information Sciences, Beijing 100193, China}
\affiliation{Hefei National Laboratory, Hefei 230088, China}
\affiliation{Beijing Key Laboratory for Advanced Quatnum Technology, Beijing 100190, China}

\author{Heng Fan}
\email{hfan@iphy.ac.cn}
\affiliation{Beijing National Laboratory for Condensed Matter Physics, Institute of Physics, Chinese Academy of Sciences, Beijing 100190, China}
\affiliation{School of Physical Sciences, University of Chinese Academy of Sciences, Beijing 100049, China}
\affiliation{Beijing Key Laboratory of Fault-Tolerant Quantum Computing, Beijing Academy of Quantum Information Sciences, Beijing 100193, China}
\affiliation{Hefei National Laboratory, Hefei 230088, China}
\affiliation{Beijing Key Laboratory for Advanced Quatnum Technology, Beijing 100190, China}

\begin{abstract}
A common challenge in superconducting quantum circuits is the trade-off between strong coupling and computational subspace integrity. We present Mathieu control, which uses a non-resonant two-photon drive to create a selective nonlinear frequency shift. This shift modifies interactions while preserving qubit states, enabling continuous tuning of the ZZ coupling, including full suppression, and integrating single- and two-qubit gates with low leakage. For a qubit-coupler-qubit device, it allows independent ZZ control, facilitating a programmable Heisenberg (XXZ) Hamiltonian. Extended to a five-qubit chain, the system can be reconfigured to simulate dynamics of quantum magnetic phases. Mathieu control thus provides a framework for high-fidelity quantum logic and programmable simulation.
\end{abstract}
\maketitle

\vspace*{-40pt}
\section{Introduction}
\vspace*{-5pt}
Superconducting circuits constitute a mature and widely studied platform for quantum information processing and simulation, where precise control of coherent interactions is fundamental. A persistent challenge in these systems is to engineer strong, tunable couplings between qubits without inducing leakage from the computational subspace.

Current strategies predominantly follow two paths: applying resonant microwave drives to mediate couplings \cite{PhysRevApplied.12.064013, PhysRevApplied.23.014062, Wei2024_nativeGates,PhysRevResearch.4.043127,PhysRevLett.129.060501}, or using static and parametric biasing of circuit elements like tunable couplers\cite{PhysRevX.11.021058, PhysRevLett.125.240503, PhysRevX.13.031035,2h4m-mg2p,PhysRevLett.125.120504}, often combined with parametric modulation \cite{PhysRevApplied.16.024050, PhysRevResearch.2.033447}. While effective in certain regimes, both approaches tie strong interaction strength to significant dressing of qubit states \cite{PhysRevX.13.031035, PhysRevX.11.021058, PhysRevResearch.2.033447}, leading to coherent errors and frequency crowding that complicate scaling. This intrinsic trade-off between tunability and subspace integrity necessitates a new control paradigm.

Here, we introduce Mathieu control, which transcends this trade-off. By applying a non-resonant, two-photon drive to the quadratic potential (e.g., a SQUID loop), we engineer a selective nonlinear frequency shift that renormalizes interactions while actively protecting the computational levels. This general principle enables, in superconducting circuits, continuous ZZ tuning through zero for a unified gate set, independent XXZ anisotropy programming, and the simulation of quantum magnetic phases in a scalable chain. By utilizing a single parametric drive for both high-fidelity logic and tunable interactions, this unification significantly simplifies the control stack.

\vspace*{-20pt}
\section{Results}
\vspace*{-10pt}
\subsection{Principle and Effective Hamiltonians}
\vspace*{-10pt}
For a frequency-tunable quantum harmonic oscillator, the Hamiltonian can be expressed in terms of generalized momentum $\hat{p}$ and coordinate $\hat{x}$ as 
\begin{eqnarray}
\hat{H}_0(t) = \frac{\hat{p}^2}{2m} + \frac{1}{2}m\omega^2(t)\hat{x}^2,
\end{eqnarray}
where $\omega(t) = \omega_0 + \Delta\omega(t)$ with $|\Delta\omega(t)| \ll \omega_0$. In superconducting qubits, $\hat{x}$ typically corresponds to the superconducting phase difference across the junction, while $\hat{p}$ represents the conjugate charge operator\cite{10.1063/1.5089550}.

Expanding around the static frequency $\omega_0$ yields
\begin{eqnarray}
\hat{H}_0(t) &&= \hbar\omega_0\hat{a}^\dagger\hat{a} + \hbar\Delta\omega(t)\left(\hat{a}^\dagger\hat{a} + \frac{1}{2} + \frac{\hat{a}^2 + \hat{a}^{\dagger2}}{2}\right) \nonumber\\
 &&~~~~+\mathcal{O}\left(\frac{\Delta\omega^2}{\omega_0}\right),
\end{eqnarray}
where $\hat{a}$ and $\hat{a}^\dagger$ are the annihilation and creation operators defined at $\omega_0$. The first term in parentheses ($\propto \hat{a}^\dagger\hat{a}+1/2$) represents the conventional longitudinal shift, routinely employed for fast frequency tuning of transmon qubits. The second term ($\propto \hat{a}^2+\hat{a}^{\dagger2}$), however, has received comparatively little attention despite being of the same magnitude.

This quadratic term enables direct two-photon transitions between Fock states of qubits, an effect has been observed in several experiments\cite{yan2016flux,liu2005a,y3qx-cs3r}. In the classical limit, the corresponding equation of motion is the Mathieu equation $\ddot{x} + \omega_0^2[1 + \epsilon\cos(2\omega_0 t)]x = 0$, which exhibits parametric resonance when driven at twice the natural frequency\cite{landau_mechanics_1976}. We therefore term this control mechanism \textbf{Mathieu control}.
\begin{figure*}
    \centering
    \includegraphics[width=1.0\linewidth]{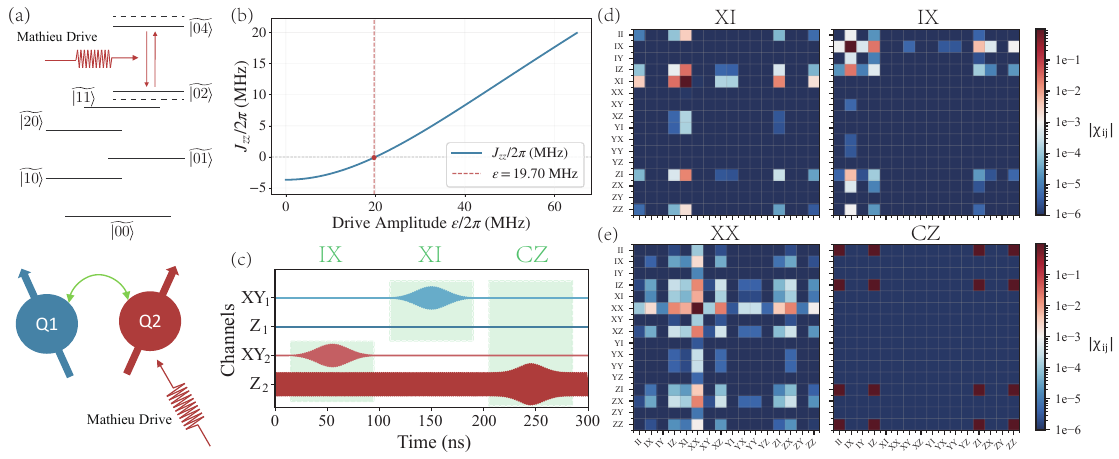}
    \captionsetup{justification=raggedright,singlelinecheck=false}
    \caption{(a) Schematic of the two directly coupled transmon qubits and the applied two-photon flux drive $\hat{H}_d(t)$ on Qubit 2. The corresponding energy level diagram illustrates the off-resonant coupling (red arrow) inducing level repulsion between states $|02\rangle$ and $|04\rangle$.
    (b) The effective ZZ coupling strength $J_{zz}$ as a function of the Mathieu control drive amplitude $\epsilon$ at a fixed drive frequency $\omega_d$, calculated by exact diagonalization of the full Hamiltonian. The vertical dashed line marks the operating point $\epsilon_0$ where $J_{zz}=0$.
    (c) Schematic of the control pulse sequence for implementing single-qubit and two-qubit gates. The sequence executes an XI gate, followed by an IX gate, and then a CZ gate. The waveforms for the XY control lines (top two panels) and the Mathieu (Z) control lines (bottom two panels) of both qubits are shown.
    (d) Process matrices ($\chi$ matrices) in the Pauli transfer representation for the simulated XI gate (left panel) and IX gate (right panel). The basis is the two-qubit Pauli basis $\{I, \sigma_x, \sigma_y, \sigma_z\}^{\otimes 2}$. Color scale represents the absolute value $|\chi_{mn}|$ on a logarithmic scale.
    (e) Process matrices ($\chi$ matrices) for the simulated simultaneous two-qubit X gate (XX, left panel) and the controlled-phase (CZ) gate (right panel). Display conventions are identical to (d).}
    \label{fig:directcouple}
\end{figure*}

Remarkably, the potential of Mathieu control extends far beyond inducing direct resonant transitions. When a high-frequency flux drive $\Delta\omega(t) = \epsilon \cos(\omega_d t)$ is applied with a deliberate detuning $\delta = \omega_d - \omega_{i,i+2}$ from a two-photon transition frequency $\omega_{i,i+2}$ (between Fock states $\ket{i}$ and $\ket{i+2}$), the interaction becomes an dispersive process. In a frame rotating at $\omega_d/2$ for the drive, the quadratic term yields, within the rotating wave approximation, an effective Hamiltonian $\hat{H}_{\text{eff}}^{(i)} \propto (\epsilon^2/\delta) \hat{a}^{\dagger 2} \hat{a}^2$. This term acts as a \textit{nonlinear frequency shift}, which induces level repulsion specifically between states $\ket{i}$ and $\ket{i+2}$ without inducing real population transfer. This mechanism provides a powerful, non-invasive ``knob'' to selectively reshape the anharmonic energy ladder. As we demonstrate next, by applying this drive to one or more elements in a coupled circuit, this capability can be harnessed to engineer widely tunable effective spin-spin interactions.

\vspace*{-20pt}
\subsection{Tunable ZZ interaction between directly coupled transmon qubits}
We first demonstrate the application of Mathieu control for engineering tunable interactions between two directly coupled transmon qubits. The system Hamiltonian is
\begin{eqnarray}
\hat{H}_0 = \sum_{i=1}^2 \left[\omega_i \hat{a}_i^\dagger \hat{a}_i + \frac{\alpha_i}{2} \hat{a}_i^\dagger \hat{a}_i^\dagger \hat{a}_i \hat{a}_i\right] + g(\hat{a}_1^\dagger \hat{a}_2 + \hat{a}_1 \hat{a}_2^\dagger),
\label{eq:H0}
\end{eqnarray}
where $\hat{a}_i$ ($\hat{a}_i^\dagger$) are the annihilation (creation) operators for qubit $i$, $\omega_i$ and $\alpha_i$ are its fundamental frequency and anharmonicity, and $g$ is the direct capacitive coupling. Applying a two-photon flux drive $\hat{H}_d(t) = \epsilon \cos(\omega_d t) (\hat{a}_2^2 + \hat{a}_2^{\dagger 2})$ to the SQUID loop of qubit 2 creates a dispersive coupling between its $\ket{2}$ and $\ket{4}$ states, as illustrated in Fig.~\ref{fig:directcouple}(a). This off-resonant process induces a nonlinear frequency shift that selectively modifies the qubit's anharmonic ladder.

A Schrieffer-Wolff transformation (detailed in Appendix D) yields an effective static Hamiltonian within the computational subspace. The dominant correction is a tunable ZZ interaction:
\begin{eqnarray}
&\hat{H}_{ZZ}^{\text{eff}} = \frac{1}{4}J_{zz} \sigma_z^1 \sigma_z^2,\\
 &J_{zz} = \frac{2g^2}{\Delta + \alpha_1} - \frac{2g^2}{(\Delta - \alpha_2) + \frac{3\epsilon^2}{\delta_d - \Delta + \alpha_2}} + \frac{3\epsilon^2}{4\alpha_2 - 2\delta_d}.\nonumber
\label{eq:Hzz_eff}
\end{eqnarray}
Here, $\Delta = \omega_2-\omega_1$ and $\delta_d = \omega_d-2\omega_2+5\alpha_2$. This analytical expression shows that $J_{zz}$ can be continuously tuned via the drive amplitude $\epsilon$, crossing zero at a specific amplitude $\epsilon_0$.

We verify this tunability by exact numerical diagonalization of the full driven Hamiltonian. The effective ZZ coupling is extracted from the exact eigenenergies $E_{ij}$ of the dressed computational states $\widetilde{|ij\rangle}$ using the standard spectroscopic definition:
\begin{eqnarray}
J_{zz} = E_{11} - E_{01} - E_{10} + E_{00}.
\label{eq:zeta_def}
\end{eqnarray}
As shown in Fig.~\ref{fig:directcouple}(b), the numerically computed $J_{zz}$ as a function of $\epsilon$ closely follows the analytical prediction, demonstrating a wide tuning range from negative to positive values, including a zero-coupling point (vertical dashed line). At this point ($J_{zz}=0$), the qubits are effectively decoupled in the computational subspace, enabling independent single-qubit operations with minimal crosstalk and establishing it as our primary operating condition.

The continuous tunability of $J_{zz}$ provides a unified framework for implementing both single- and two-qubit gates within the same circuit. Figure~\ref{fig:directcouple}(c) illustrates the control sequence. When the drive amplitude is set to $\epsilon_0$ (where $J_{zz}=0$), the two qubits are effectively decoupled, allowing for independent single-qubit operations. We perform single-qubit X rotations (XI, IX) and a simultaneous two-qubit X (XX) gate by applying standard Gaussian-enveloped microwave pulses to the XY control lines. To implement a controlled-phase (CZ) gate, we adiabatically increase $\epsilon$ away from $\epsilon_0$ to induce a finite $J_{zz}$\cite{PhysRevA.90.022307}, which imparts a conditional phase to the $\ket{11}$ state.

We characterize these gates via quantum process tomography, extracting the process matrices $\chi$ in the Pauli transfer representation (Figs. \ref{fig:directcouple}(d) and (e)). The $\chi$ matrices closely match their ideal targets. The achieved process fidelities $F_{\chi}$ and the corresponding $\text{Tr}(\chi^2)$ values, which quantify the preservation of unitarity, are summarized below:

\begin{center}
\begin{tabular}{lcc}
\hline
\textbf{Gate} & $F_{\chi}$ (\%) & $\text{Tr}(\chi^2)$ \\
\hline
XI & 99.945 & 0.9999951 \\
IX & 99.970 & 0.99999987 \\
XX & 99.947 & 0.999966 \\
CZ & 99.994 & 0.999935 \\
\hline
\end{tabular}
\end{center}

While the single-qubit gate fidelities have room for improvement—primarily due to residual crosstalk, a key insight comes from the $\text{Tr}(\chi^2)$ values. For all gates, $1 - \text{Tr}(\chi^2) \sim 10^{-5}$, confirming that the error process is overwhelmingly unitary and confined to the computational subspace. This is a direct signature of the \textbf{low leakage} inherent to the Mathieu control scheme: population transfer to non-computational states remains below the $10^{-5}$ level. This error structure is favorable for fault-tolerant quantum computing, as unitary errors within the computational subspace can often be mitigated via compilation algorithms\cite{PhysRevX.11.041039,gambetta2017building,blume2017demonstration}, whereas leakage errors pose a far more severe challenge.
\begin{figure*}
    \centering
    \includegraphics[width=1.0\linewidth]{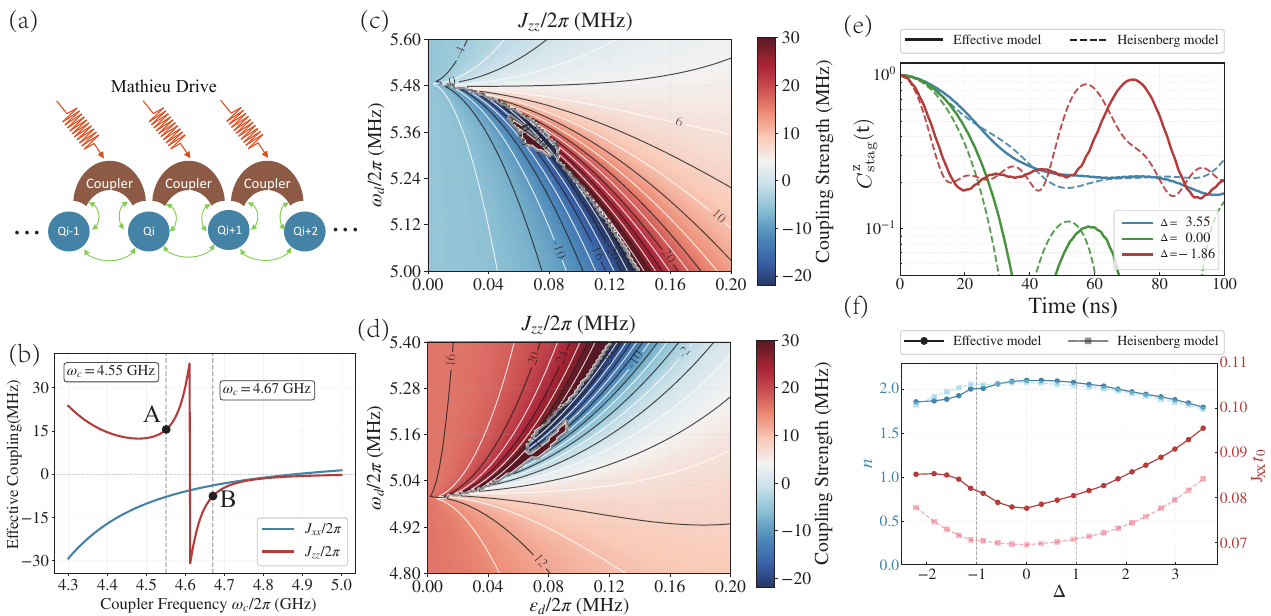}
    \captionsetup{justification=raggedright,singlelinecheck=false}
    \caption{(a) Schematic of an extensible and programmable Heisenberg interaction chain realized with an alternating arrangement of qubits and tunable couplers. Green arrows indicate the static capacitive coupling strengths between adjacent elements. Red arrows indicate the application of Mathieu control pulses exclusively to the couplers' SQUID loops, while the qubits remain idle.
    (b) Static tuning of the effective qubit-qubit couplings $J_{xx}$ and $J_{zz}$ in a single qubit-coupler-qubit unit as a function of the coupler frequency $\omega_c$. Working points A ($\omega_c = 4.55\,\text{GHz}$) and B ($\omega_c = 4.67\,\text{GHz}$) are marked.
    (c) Two-dimensional map of the effective $J_{zz}$ coupling strength for the system biased at static working point A under the applied two-photon Mathieu drive. The coupling is plotted as a function of the drive amplitude $\epsilon$ and frequency $\omega_d$. Contour lines are spaced at intervals of $2\,\text{MHz}$.
    (d) Corresponding two-dimensional map of the effective $J_{zz}$ coupling for the system biased at static working point B.
    (e) Normalized staggered $\sigma^z$ correlation function $\mathcal{C}^z_{\text{stag}}(t)$ as a function of time for three representative values of the anisotropy parameter $\Delta = J_{zz}/2J_{xx}$: $3.55$, $0.00$, and $-1.86$. The solid lines show the evolution in the five-qubit chain under Mathieu control. Dashed lines show an exact five-qubit Heisenberg chain.
    (f) Fitted parameters $t_0$ (right axes) and $n$ (left axes) obtained by fitting the early-time relaxation of $\mathcal{C}^z_{\text{stag}}(t)$ to the function $\exp[-(t/t_0)^n]$. The solid lines correspond to the Mathieu-controlled chain at different efficient $\Delta$ values. Light dashed lines show the corresponding fits for the exact five-qubit Heisenberg chain.} 
    \label{fig:qcq_scheme}
\end{figure*}

\subsection{Scalable Engineering of Programmable Heisenberg Interactions}
We now demonstrate that Mathieu control provides a scalable path to engineer fully programmable Heisenberg-type interactions. We first analyze the basic building block: a three-element chain consisting of two computational transmon qubits coupled via a tunable transmon coupler (the QCQ architecture), as illustrated in Fig.~\ref{fig:qcq_scheme}(a). The static system Hamiltonian is
\begin{eqnarray}
\hat{H}_0 = && \sum_{i\in\{1,2,c\}} \left( \omega_i a_i^\dagger a_i + \frac{\alpha_i}{2} a_i^\dagger a_i^\dagger a_i a_i \right) \nonumber\\
&& + \sum_{i\ne j} g_{ij} (a_i^\dagger a_j + a_i a_j^\dagger),
\label{eq:H0_QCQ}
\end{eqnarray}
where the operators $a_c$ ($a_c^\dagger$) act on the coupler with frequency $\omega_c$ and anharmonicity $\alpha_c$. The couplings $g_{ic}$ connect qubit $i$ to the coupler, while $g_{12}$ denotes a small direct capacitive coupling between the qubits. We focus on the resonant case $\omega_1 = \omega_2 \ne \omega_c$. In the dispersive regime ($|\omega_c - \omega_{1,2}| \gg g_{ic}$), the effective qubit-qubit Hamiltonian within the computational subspace is well-approximated by an anisotropic Heisenberg (XXZ) model:
\begin{eqnarray}
H_{\text{eff}} = \frac{J_{xx}}{2} (\sigma_x^1 \sigma_x^2 + \sigma_y^1 \sigma_y^2) + \frac{J_{zz}}{4} \sigma_z^1 \sigma_z^2,
\label{eq:H_eff_undriven}
\end{eqnarray}
where the coupling strengths are extracted from the low-energy spectrum: $J_{xx} = (E_{100} - E_{001})/2$ and $J_{zz} = E_{101} - E_{100} - E_{001} + E_{000}$.

Static tuning of the coupler frequency $\omega_c$ modifies both $J_{xx}$ and $J_{zz}$ simultaneously, as shown in Fig.~\ref{fig:qcq_scheme}(b). While this provides a basic control knob, it does not allow for independent adjustment of the XY and Ising interaction strengths, which is crucial for simulating a wide range of quantum spin models.

Mathieu control overcomes this limitation. Applying a two-photon drive selectively to the coupler's SQUID loop,
\begin{eqnarray}
    \hat{H}_d(t) = \epsilon \cos(\omega_d t) \left( a_c^2 + a_c^{\dagger 2} \right),
    \label{eq:Hdrive_QCQ}
\end{eqnarray}
introduces a powerful and largely independent control over the ZZ interaction. The drive resonantly couples the coupler's $|2\rangle_c$ and $|4\rangle_c$ states, dressing its energy levels and thereby modifying the dispersive shifts that determine $J_{zz}$. We investigate two static working points, A ($\omega_c = 4.55\,\text{GHz}$) and B ($\omega_c = 4.67\,\text{GHz}$), marked in Fig.~\ref{fig:qcq_scheme}(b). As shown in Figs.~\ref{fig:qcq_scheme}(c) and (d), sweeping the drive amplitude $\epsilon$ and frequency $\omega_d$ enables continuous, wide-ranging tuning of $J_{zz}$—from strong suppression to significant enhancement—while leaving $J_{xx}$ virtually unchanged. This demonstrates the key feature of selective ZZ control, decoupling it from the XY exchange process. The detailed analytical mechanism is provided in Appendix~E.

To showcase the scalability and programmability of this approach, we extend it to a five-qubit chain with alternating qubits and couplers [Fig.~\ref{fig:qcq_scheme}(a)]. All couplers are first biased to a common static working point (Point B), establishing a baseline uniform XXZ interaction. Mathieu control pulses are then applied simultaneously to all couplers, with identical amplitude $\epsilon_d$ and frequency $\omega_d$, to uniformly reprogram the anisotropy parameter $\Delta \equiv J_{zz}/(2J_{xx})$ across the entire chain.

We probe the simulated quantum magnetism by studying the relaxation dynamics of a dressed Néel initial state, $|\Psi_0\rangle = \ket{\uparrow\downarrow\uparrow\downarrow\uparrow}_{\text{dressed}}$(See APPENDIX C.4), which accounts for the residual dressing from the nearby couplers. We monitor the normalized staggered $\sigma^z$ correlation function,
\begin{eqnarray}
\mathcal{C}^z_{\text{stag}}(t) = \langle(-1)^{|i-j|}\hat{\sigma}_z^i\hat{\sigma}_z^j \rangle_{i,j},
\end{eqnarray}
where $\langle\cdot\rangle_{i,j}$ denotes averaging over all $i\ne j$. Figure~\ref{fig:qcq_scheme}(e) shows $\mathcal{C}^z_{\text{stag}}(t)$ for three representative values of the programmed anisotropy $\Delta$, covering the gapless XY ($|\Delta|<1$), gapped Ising-antiferromagnetic ($\Delta>1$), and ferromagnetic ($\Delta<-1$) regimes. The dashed lines show the evolution under an exact five-qubit Heisenberg chain with the same nominal couplings.

The agreement between our Mathieu-controlled chain and the exact model is excellent in the XY and Ising-antiferromagnetic phases. The decay curves match closely in functional form. The only systematic deviation is a uniform slowdown in relaxation—the characteristic decay time $t_0$ in our system is approximately one-sixth longer. This slowdown has a physical origin: in the extended chain, the effective spin-spin coupling is renormalized and suppressed due to (i) increased connectivity (inner qubits couple to two couplers) and (ii) dressing of the computational basis, collectively reducing the overall energy scale of the simulated Hamiltonian.

A more comprehensive analysis across a wide range of $\Delta$ is presented in Fig.~\ref{fig:qcq_scheme}(f). We fit the early-time relaxation to a stretched exponential $\exp[-(t/t_0)^n]$. In the $|\Delta|<1$ and $\Delta>1$ regimes, the extracted exponent $n$ matches the exact model well, confirming the accuracy of the simulated relaxation dynamics. Most notably, the evolution of $t_0$ clearly signals the quantum phase transitions at $\Delta =  1$: as $\Delta$ increases through 1, the trend in $t_0$ changes, marking the transition from the gapless XY phase to the gapped Ising-antiferromagnetic phase\cite{giamarchi2003quantum, PhysRevB.12.3908, lukin2019probing}.

In the ferromagnetic regime ($\Delta < -1$), the correlation decay is strongly suppressed and exhibits pronounced oscillations—a characteristic feature of finite-size chains in the ferromagnetic phase. While the overall relaxation trend remains consistent (Fig.~\ref{fig:qcq_scheme}(e)), the stretched-exponential fit becomes less accurate due to these oscillations, leading to larger uncertainties in the fitted parameters $t_0$ and $n$ (Fig.~\ref{fig:qcq_scheme}(f)). This is expected for small systems where finite-size effects are significant, and the simple fitting function may not fully capture the complex dynamics.

The successful reproduction of the XXZ phase diagram—including the critical behavior at the phase boundaries—through nonequilibrium dynamics underscores the high programmability and simulation fidelity achieved by Mathieu control in a scalable architecture.

\vspace*{-10pt}
\section{Conclusion}
\vspace*{-10pt}
We have demonstrated Mathieu control as a paradigm that decouples interaction strength from computational subspace dressing. By exploiting a non-resonant two-photon nonlinearity, it enables both a unified gate set in fixed-frequency architectures and, crucially, independent in-situ programming of Heisenberg interactions.

The underlying principle—selective spectral engineering via off-resonant parametric drives—is general and could be applied to other platforms featuring tunable oscillators, such as trapped ions or nanomechanical systems. Our work thus provides a powerful tool for high-fidelity quantum logic and opens a path for programmable simulation of complex Hamiltonians.

\begin{acknowledgments}
We thank the support from the Synergetic Extreme Condition User Facility (SECUF) in Huairou District, Beijing. This work was supported by the National Natural Science Foundation of China (Grants No. 92265207, No. T2121001, No. U25A6009, No. T2322030, No. 12122504, No. 12274142, and No. 12475017), the Innovation Program for Quantum Science and Technology (Grant No. 2021ZD0301800), the Beijing Nova Program (Grant No. 20220484121), and the Ministry of Science and Technology project (Grant No. 2025YFE0217600).
\end{acknowledgments}

\newpage
\appendix
\section*{Appendix A: Rotating Frame Transformation and Hamiltonian Engineering}

In this section, we derive the effective Hamiltonian in the rotating frame used to simulate the adiabatic evolution and leakage analysis presented in the main text. The numerical implementation performs a unitary transformation to remove the explicit time dependence of the drive, enabling the calculation of quasi-static eigenenergies and eigenstates.

\subsection*{A.1 The Lab Frame Hamiltonian}

We model the system as two coupled weakly anharmonic oscillators (transmons). In the laboratory frame, the static Hamiltonian $\hat{H}_{\text{lab}}$ (setting $\hbar = 1$) is given by:

\begin{eqnarray}
    \hat{H}_{\text{lab}} = \sum_{k=1,2} \left( \omega_k \hat{n}_k - \frac{\alpha_k}{2} \hat{a}_k^\dagger \hat{a}_k^\dagger \hat{a}_k \hat{a}_k \right) + g (\hat{a}_1^\dagger \hat{a}_2 + \hat{a}_1 \hat{a}_2^\dagger),\nonumber\\
\end{eqnarray}

\noindent where $\omega_k$ is the bare frequency of qubit $k$, $\alpha_k$ is the anharmonicity, $\hat{a}_k$ ($\hat{a}_k^\dagger$) are the annihilation (creation) operators, and $\hat{n}_k = \hat{a}_k^\dagger \hat{a}_k$ is the number operator. The coupling strength between the qubits is denoted by $g$.

\subsection*{A.2 Parametric Drive and The Rotating Frame}

To analyze the system dynamics under a parametric drive with frequency $\omega_d$ (where $\omega_d \approx 2\omega_k$ or $\omega_d \approx \omega_1 + \omega_2$), we transform the system into a frame rotating at half the drive frequency. The unitary transformation operator is defined as:

\begin{eqnarray}
    \hat{U}(t) = \exp\left[ -i \frac{\omega_d}{2} (\hat{n}_1 + \hat{n}_2) t \right].
\end{eqnarray}

\noindent The effective Hamiltonian in the rotating frame, $\tilde{H}$, is obtained via the transformation:

\begin{eqnarray}
    \tilde{H} = \hat{U}^\dagger(t) \hat{H}_{\text{lab}} \hat{U}(t) - i \hat{U}^\dagger(t) \frac{d\hat{U}}{dt}.
\end{eqnarray}

\noindent The second term (the geometric correction) yields $-\frac{\omega_d}{2}(\hat{n}_1 + \hat{n}_2)$. This corresponds directly to our computational implementation:
\begin{eqnarray}
\tilde{H} = \hat{H}_{lab} - \frac{\omega_d}{2}(\hat{n}_1 + \hat{n}_2).
\end{eqnarray}

\subsection*{A.3 Rotating Wave Approximation (RWA)}

Under this transformation, the number operators $\hat{n}_k$ and the anharmonic terms $\hat{a}_k^{\dagger 2} \hat{a}_k^2$ remain invariant. However, the external drive terms, which typically take the form of voltage drives $\propto \cos(\omega_d t + \phi)$, interact with the non-linear operators.

For a parametric drive intended to activate two-photon transitions (e.g., squeezing or $|0\rangle \leftrightarrow |2\rangle$ transitions), the relevant interaction in the rotating frame involves terms that become static (time-independent).

To derive the specific form of the drive term, we consider a physical parametric drive in the laboratory frame acting on the system. The drive Hamiltonian is typically modeled as a modulation of the potential with amplitude $A_X$ and frequency $\omega_d$, coupling to the quadrature operator squared:

\begin{eqnarray}
    \hat{H}_{\text{drive}}(t) = A_X \cos(\omega_d t + \phi) (\hat{a} + \hat{a}^\dagger)^2.
\end{eqnarray}

\noindent Upon transforming to the rotating frame via the unitary operator $\hat{U}(t) = e^{-i \frac{\omega_d}{2} \hat{n} t}$, the annihilation operator transforms as $\hat{a} \to \hat{a} e^{-i \frac{\omega_d}{2} t}$. Substituting this into the drive Hamiltonian yields:

\begin{eqnarray}
    \tilde{H}_{\text{drive}}&&=\hat{U}^\dagger \hat{H}_{\text{drive}}(t) \hat{U} \nonumber\\
    &&=A_X \left[ \frac{e^{i(\omega_d t + \phi)} + e^{-i(\omega_d t + \phi)}}{2} \right] \nonumber\\
    &&~~~~\times\left( \hat{a} e^{-i \frac{\omega_d}{2} t} + \hat{a}^\dagger e^{i \frac{\omega_d}{2} t} \right)^2.\nonumber\\
\end{eqnarray}

\noindent Expanding the squared term, we obtain terms oscillating at different frequencies:
\begin{itemize}
    \item $\hat{a}^2 e^{-i \omega_d t}$
    \item $(\hat{a}^\dagger)^2 e^{i \omega_d t}$
    \item $\hat{a}^\dagger \hat{a} + \hat{a} \hat{a}^\dagger$ (constant in time)
\end{itemize}

\noindent Multiplying these by the cosine factor $\cos(\omega_d t + \phi)$, we isolate the terms that become stationary (time-independent). The condition for a term to be static is that its time-dependent phase factors cancel out. Specifically:
\begin{itemize}
    \item The $\hat{a}^2 e^{-i \omega_d t}$ term couples with $e^{i(\omega_d t + \phi)}$, yielding $\hat{a}^2 e^{i\phi}$.
    \item The $(\hat{a}^\dagger)^2 e^{i \omega_d t}$ term couples with $e^{-i(\omega_d t + \phi)}$, yielding $(\hat{a}^\dagger)^2 e^{-i\phi}$.
\end{itemize}
Terms oscillating at $2\omega_d$ or $\omega_d$ are discarded under the RWA. Applying results in the effective static drive Hamiltonian:

\begin{eqnarray}
    \tilde{H}_{\text{drive}} = \sum_{k=1,2} \left( \frac{\Omega_k}{2} \hat{a}_k^2 e^{i \phi_k} + \frac{\Omega_k^*}{2} (\hat{a}_k^\dagger)^2 e^{-i \phi_k} \right).
\end{eqnarray}

\noindent Here, $\Omega_k$ represents the effective two-photon Rabi frequency (proportional to the drive amplitude $A_X$).

\subsection*{A.4 Hamiltonian for Adiabatic Leakage Analysis}

Combining the transformed static terms and the RWA drive, the total effective Hamiltonian used for spectral analysis is:

\begin{eqnarray}
    \tilde{H}_{\text{RWA}} = \sum_{k=1,2} \left( \Delta_k \hat{n}_k - \frac{\alpha_k}{2} \hat{a}_k^\dagger \hat{a}_k^\dagger \hat{a}_k \hat{a}_k \right) + \tilde{H}_{\text{int}} + \tilde{H}_{\text{drive}},\nonumber\\
    ~~
\end{eqnarray}

\noindent where the detuning is defined as $\Delta_k = \omega_k - \omega_d/2$. 

This time-independent Hamiltonian $\tilde{H}_{\text{RWA}}$ allows us to calculate the quasi-energy spectrum. In the simulation, we diagonalize this matrix to find the instantaneous eigenstates $|\psi_j\rangle$. The adiabatic leakage is then quantified by projecting the time-evolved state onto the instantaneous eigenstates of the driven system, determining transitions to non-computational subspaces due to finite-time ramp speeds.

\subsection*{B.1 Mathieu Control Hamiltonian}
As derived in Appendix A, the Mathieu control is simulated in a frame rotating at $\omega_d/2$. The effective Hamiltonian is:
\begin{equation}
    \hat{H}_{\text{Mathieu}} = \hat{H}_{\text{static}} + \frac{\Omega_2(t)}{2}\hat{a}_2^2 + \frac{\Omega_2^*(t)}{2}(\hat{a}_2^\dagger)^2,
\end{equation}
where $\Omega_2(t)$ is the complex Rabi frequency proportional to the drive amplitude.

\subsection*{B.2 Parameters for Mathieu Control (Fig.~\ref{fig:directcouple})}
The system parameters for the directly coupled transmon qubits under Mathieu control are listed in Table \ref{tab:mathieu_params}. 

\begin{table}[h]
    \centering
    \caption{System and drive parameters for the Mathieu control simulations.}
    \label{tab:mathieu_params}
    \begin{tabular}{lcc}
        \hline
        \textbf{Physical Parameters} & \textbf{Symbol} & \textbf{Value (GHz)} \\
        \hline
        Qubit 1 Frequency & $\omega_1/2\pi$ & $5.20$ \\
        Qubit 2 Frequency & $\omega_2/2\pi$ & $5.75$ \\
        Anharmonicity & $\alpha_{1,2}/2\pi$ & $0.25$ \\
        Coupling Strength & $g/2\pi$ & $0.03$ \\
        \hline
        \textbf{Drive Settings} & & \\
        \hline
        Fig.~\ref{fig:directcouple}(b) Drive Freq. & $\omega_d^{(b)}/2\pi$ & $10.60$ \\
        \hline
    \end{tabular}
\end{table}

\subsection*{B.3 Adiabatic Pulse Optimization}

To realize high-fidelity adiabatic gates, we employ a pulse shaping technique that dynamically adjusts the drive ramp rate based on the instantaneous leakage risk:

\textbf{1. Leakage Sensitivity Profiling:}
First, we diagonalize the instantaneous Hamiltonian $\hat{H}(\mathcal{E})$ over the full range of drive amplitudes $\mathcal{E}$. For the target computational state (e.g., $|11\rangle$), we define a leakage sensitivity function $S(\mathcal{E})$ and an energy gap function $\Delta E(\mathcal{E})$:
\begin{equation}
\begin{aligned}
    S(\mathcal{E}) &= \sum_{k \in \text{leak}} \left| \frac{\langle \psi_k | \hat{H}_{\text{leak}} | \psi_{\text{target}} \rangle}{E_k - E_{\text{target}}} \right| \\
    \Delta E(\mathcal{E}) &= \min_{k} |E_k - E_{\text{target}}|
\end{aligned}
\end{equation}
where $\hat{H}_{\text{leak}}$ represents the drive operator connecting the subspace.

\textbf{2. ODE-Based Waveform Generation:}
The optimal pulse waveform $\mathcal{E}(t)$ is generated by solving a coupled ordinary differential equation (ODE). We seek a parametrization that limits the diabatic transitions. The ramp rate $d\mathcal{E}/dt$ is constrained by the adiabatic condition:
\begin{equation}
    \frac{d\mathcal{E}}{dt} = k \cdot \frac{\Delta E(\mathcal{E})}{S(\mathcal{E})} \cdot f(\tau),
\end{equation}
where $k$ is a scaling factor controlling the total gate time, and $f(\tau)$ is a normalized window function (e.g., sine-squared) to ensure smooth boundary conditions.
\subsection*{B.4 XY Control Hamiltonian}

To demonstrate single-qubit control (e.g., X gate) within our architecture, we introduce an XY control line. The drive field in the laboratory frame is described by:
\begin{equation}
    \hat{H}_{\text{XY, lab}}(t) = f(t) \cos(\omega_{xy} t) (\hat{a}^\dagger + \hat{a}),
\end{equation}
where $f(t)$ is the pulse envelope and $\omega_{xy}$ is the carrier frequency of the XY drive.

Since our system simulation is performed in a frame rotating at $\omega_d/2$, we must transform this drive Hamiltonian accordingly. The transformation operator is $\hat{U}(t) = \exp[-i (\omega_d/2) t \hat{n}]$. Applying the rotating wave approximation (RWA) to discard the fast-oscillating terms (at frequencies $\pm(\omega_{xy} + \omega_d/2)$), the effective Hamiltonian for the XY channel becomes:
\begin{equation}
    \tilde{H}_{\text{XY}} = \frac{f(t)}{2} \left( \hat{a}^\dagger e^{-i(\omega_{xy} - \omega_d/2)t} + \hat{a} e^{i(\omega_{xy} - \omega_d/2)t} \right).
\end{equation}
This Hamiltonian governs the single-qubit rotations demonstrated in the main text. The detuning term $(\omega_{xy} - \omega_d/2)$ accounts for the frequency difference between the XY drive and the rotating frame of the parametric coupler.

\section*{Appendix C: Tunable Coupler Architecture (QCQ) Parameters}

In this section, we describe the Hamiltonian and the parameter regime for the Quantum-Coupler-Quantum (QCQ) architecture. This setup utilizes a tunable nonlinear coupler to mediate effective interactions between two fixed-frequency qubits.

\subsection*{C.1 Hamiltonian in the Rotating Frame}

The system comprises two transmon qubits (Q1, Q2) and a central coupler (C). The total Hamiltonian in the laboratory frame includes the static coupled terms and the parametric drive. To analyze the effective dynamics, we move to a rotating frame where the explicit time-dependence of the drive is removed, similar to the transformation discussed in Appendix A.

The effective static Hamiltonian in this frame is given by:
\begin{eqnarray}
    \hat{H}_{\text{eff}} = \hat{H}_{\text{sys}} + \hat{H}_{\text{drive}}.
\end{eqnarray}

The system Hamiltonian $\hat{H}_{\text{sys}}$ describes three coupled anharmonic oscillators:
\begin{eqnarray}
    \hat{H}_{\text{sys}} &&= \sum_{j \in \{1, c, 2\}} \left[ \Delta_j \hat{n}_j - \frac{\alpha_j}{2} (\hat{a}_j^\dagger)^2 \hat{a}_j^2 \right] \nonumber\\
    &&~~~~+ \sum_{\langle i, j \rangle} g_{ij} (\hat{a}_i^\dagger \hat{a}_j + \hat{a}_i \hat{a}_j^\dagger).
\end{eqnarray}
Here, $\Delta_j$ represents the detuning in the rotating frame. The coupling is dominated by the qubit-coupler terms ($g_{1c}, g_{2c}$), with a residual direct qubit-qubit coupling $g_{12}$.

To activate the parametric exchange (XX) interaction, a two-photon drive is applied exclusively to the coupler mode. In the appropriate rotating frame (where the frame frequency matches half the drive frequency for the coupler mode), this drive becomes static:
\begin{eqnarray}
    \hat{H}_{\text{drive}} = \pi \mathcal{E}_c (\hat{a}_c^2 + \hat{a}_c^{\dagger 2}),
\end{eqnarray}
where $\mathcal{E}_c$ is the drive amplitude. This term mixes the computational states via the coupler's higher levels, generating an effective transverse coupling $J_{\text{eff}}$ between Q1 and Q2.

\subsection*{C.2 Extraction of Effective Couplings}

We numerically diagonalize the effective Hamiltonian $\hat{H}_{\text{eff}}$ to obtain its eigenenergies. The effective interaction rates are extracted directly from the energy spectrum of the computational basis states $|q_1, c, q_2\rangle$:

\textbf{1. Effective XX Coupling ($J_{\text{eff}}$):}
The parametric drive induces a hybridization between the states $|100\rangle$ and $|001\rangle$. The effective exchange coupling strength is determined by the avoided crossing splitting:
\begin{eqnarray}
    J_{\text{eff}} = \frac{1}{2} \left| E_{001} - E_{100} \right|.
\end{eqnarray}

\textbf{2. Residual ZZ Interaction ($\chi_{ZZ}$):}
The unwanted static and dynamic ZZ crosstalk is quantified by the non-linearity of the computational energy levels:
\begin{eqnarray}
    \chi_{ZZ} = (E_{101} - E_{100}) - (E_{001} - E_{000}).
\end{eqnarray}
By scanning the drive frequency and amplitude, we identify operating points where $J_{\text{eff}}$ is maximized while $\chi_{ZZ}$ is suppressed.

\subsection*{C.3 Simulation Parameters}

The parameters used for the QCQ parameter sweep are listed in Table \ref{tab:qcq_params}. A key feature of this configuration is the large anharmonicity of the coupler ($\alpha_c = 800$ MHz), which is significantly higher than that of typical transmons. While this regime (resembling a heavy fluxonium or highly shunted mode) helps minimize leakage to non-computational states, its primary purpose is to enable precise level engineering. By detuning the coupler far from the qubits, this large anharmonicity ensures that the coupler's $|2\rangle$ state remains energetically close to the qubits' $|101\rangle$ computational state. This proximity is crucial for inducing a strong dispersive shift via off-resonant coupling, which in turn allows for wide-range tuning of the effective $J_{zz}$ interaction, as demonstrated in the main text.

\begin{table}[h]
    \centering
    \caption{Parameters for the QCQ architecture simulation. The coupler features a large anharmonicity to support high-fidelity parametric operations.}
    \label{tab:qcq_params}
    \begin{tabular}{lcc}
        \hline
        \textbf{Parameter} & \textbf{Symbol} & \textbf{Value (GHz)} \\
        \hline
        Qubit 1 Frequency & $\omega_1/2\pi$ & $4.20$ \\
        Qubit 2 Frequency & $\omega_2/2\pi$ & $4.20$ \\
        Coupler Frequency & $\omega_c/2\pi$ & Variable \\
        \hline
        Qubit Anharmonicity & $\alpha_{1,2}/2\pi$ & $0.20$ \\
        \textbf{Coupler Anharmonicity} & $\boldsymbol{\alpha_c/2\pi}$ & $\mathbf{0.80}$ \\
        \hline
        Qubit-Coupler Coupling & $g_{1c}, g_{2c}/2\pi$ & $0.08$ \\
        Direct Coupling & $g_{12}/2\pi$ & $0.01$ \\
        \hline
    \end{tabular}
\end{table}

\subsection*{C.4 Preparation of the Dressed Néel State}
\label{sec:dressed-Neel}

To simulate the dynamics of the Heisenberg model, the system is initialized in a Néel state (e.g., $|\dots 0101 \dots\rangle$). In our architecture, the presence of couplers introduces non-negligible dressing to the qubit states. Consequently, simply initializing the system in a bare product state would result in low fidelity with respect to the true eigenstates of the static Hamiltonian. Instead, we construct a "dressed" Néel state by performing a projection based on the eigenstates of local subsystems.

Specifically, for a chain of five qubits and four couplers, we construct the initial state by dividing the system into segments. Taking the left segment consisting of four devices (Qubit-Coupler-Qubit-Coupler) as an example, we first obtain the eigenstates of this subsystem. We identify the dressed eigenstates corresponding to the relevant computational basis states, denoted as $\tilde{v}_1 = \widetilde{|1000\rangle}$ and $\tilde{v}_2 = \widetilde{|0010\rangle}$. We then define a projection operator onto this subspace:
\begin{equation}
    \hat{P} = \tilde{v}_1 \tilde{v}_1^\dagger + \tilde{v}_2 \tilde{v}_2^\dagger.
\end{equation}
The local initial state for this segment, $v_{\text{left}}$, is obtained by projecting the target bare product state (e.g., $|0010\rangle$) into this dressed subspace:
\begin{equation}
    v_{\text{left}} = \hat{P} |0010\rangle_{\text{bare}}.
\end{equation}
A similar procedure is applied to the rightmost four devices to obtain $v_{\text{right}}$. The central qubit is initialized in its corresponding bare Fock state, $v_{\text{center}}$. The total initial dressed Néel state is finally formed by the tensor product of these components:
\begin{equation}
    |\Psi_{\text{init}}\rangle = v_{\text{left}} \otimes v_{\text{center}} \otimes v_{\text{right}}.
\end{equation}
This approach ensures that the simulation begins in a state that properly incorporates the static couplings introduced by the intermediate couplers.

\section{APPENDIX D: Schrieffer-Wolff Analysis of ZZ Coupling Control}
\label{sec:schrieffer_wolff}

We consider two directly coupled transmon qubits, where the second qubit is driven by a parametric drive. After moving to a rotating frame at frequency $\omega_d/2$, the Hamiltonian is given by:

\begin{eqnarray}
\hat{H} = \hat{H}_1 + \hat{H}_2 + \hat{H}_d + \hat{V},
\end{eqnarray}

where
\begin{eqnarray}
\hat{H}_1 &&= (\omega_1 - \omega_d/2) \hat{a}_1^\dagger\hat{a}_1 - \frac{\alpha_1}{2} \hat{a}_1^{\dagger 2}\hat{a}_1^2, \label{eq:H1} \nonumber\\
\hat{H}_2 &&= (\omega_2 - \omega_d/2) \hat{a}_2^\dagger\hat{a}_2 - \frac{\alpha_2}{2} \hat{a}_2^{\dagger 2}\hat{a}_2^2, \label{eq:H2} \nonumber\\
\hat{H}_d &&= \frac{\epsilon}{2}(\hat{a}_2^{\dagger 2} + \hat{a}_2^2), \label{eq:Hd} \nonumber\\
\hat{V} &&= g(\hat{a}_1^\dagger\hat{a}_2 + \hat{a}_2^\dagger\hat{a}_1), \label{eq:V}
\end{eqnarray}

The drive frequency is chosen as $\omega_d = 2\omega_2 - 5\alpha_2 + \delta_d$, where $\delta_d$ is a small detuning comparable to $\epsilon$ and much smaller than $2\alpha_2$. This choice resonantly couples the $|2\rangle$ and $|4\rangle$ states of the second qubit.

We first diagonalize $\hat{H}_2 + \hat{H}_d$ to obtain the dressed states and energies of the second qubit. The drive strongly couples $|2\rangle$ and $|4\rangle$ while other states are only weakly perturbed. To first order in $\epsilon$, the dressed eigenstates and energies are:

\begin{eqnarray}
|E_0\rangle_2 &&= |0\rangle_2, \quad E^2_0 = -\frac{\epsilon^2(2\alpha_2-\delta_d)}{(4\alpha_2-3\delta_d/2)^2-r^2}, \nonumber\\
|E_1\rangle_2 &&= |1\rangle_2, \quad E^2_1 = \frac{5\alpha_2}{2}-\frac{\delta_d}{2}-\frac{3\epsilon^2}{4\alpha_2-2\delta_d}, \nonumber\\
|E_2\rangle_2 &&= \sqrt{\frac{1}{2}\left(1+\frac{\delta_d}{2r}\right)}|4\rangle_2 + \sqrt{\frac{1}{2}\left(1-\frac{\delta_d}{2r}\right)}|2\rangle_2,\nonumber\\
E^2_2 &&= 4\alpha_2-\frac{3\delta_d}{2}-r, \nonumber\\
|E_3\rangle_2 &&= \sqrt{\frac{1}{2}\left(1+\frac{\delta_d}{2r}\right)}|2\rangle_2 - \sqrt{\frac{1}{2}\left(1-\frac{\delta_d}{2r}\right)}|4\rangle_2, \nonumber\\ 
E^2_3 &&= 4\alpha_2-\frac{3\delta_d}{2}+r,
\end{eqnarray}

where $r = \sqrt{\delta_d^2/4 + 3\epsilon^2}$.

For the first qubit, the eigenstates remain Fock states with energies:
\begin{eqnarray}
E^1_n &&= n\left(\frac{5\alpha_2}{2} - \Delta - \frac{\delta_d}{2}\right) - \frac{\alpha_1}{2}n(n-1), \nonumber\\
\Delta &&= \omega_2 - \omega_1.
\end{eqnarray}

The computational subspace is spanned by the product states $|ij\rangle = |E_i\rangle_1 \otimes |E_j\rangle_2$ for $i,j \in \{0,1\}$, with unperturbed energies:
\begin{eqnarray}
E_{00} &&= E^1_0 + E^2_0, \quad E_{01} = E^1_0 + E^2_1, \nonumber\\
E_{10} &&= E^1_1 + E^2_0, \quad E_{11} = E^1_1 + E^2_1.
\end{eqnarray}

We now apply a Schrieffer-Wolff transformation to eliminate the coupling $\hat{V}$ between the computational subspace $\mathcal{P}$ and higher energy levels $\mathcal{Q}$. The generator $S$ is chosen such that:
\begin{eqnarray}
[S, \hat{H}_0] = -\hat{V}_{\text{od}},
\end{eqnarray}
where $\hat{H}_0 = \hat{H}_1 + \hat{H}_2 + \hat{H}_d$ and $\hat{V}_{\text{od}}$ denotes the off-diagonal part of $\hat{V}$ connecting $\mathcal{P}$ and $\mathcal{Q}$.

To second order, the effective Hamiltonian in $\mathcal{P}$ is:
\begin{eqnarray}
\hat{H}_{\text{eff}} = \hat{H}_{0,\mathcal{PP}} + \frac{1}{2} [S, \hat{V}_{\text{od}}]_{\mathcal{PP}}.
\end{eqnarray}

The matrix elements of $S$ are:
\begin{eqnarray}
\langle m| S |n \rangle = \frac{\langle m| \hat{V} |n \rangle}{E_m - E_n}, \quad |m\rangle \in \mathcal{P}, |n\rangle \in \mathcal{Q}.
\end{eqnarray}

The dominant contributions to the energy shifts come from virtual transitions to states outside $\mathcal{P}$. For states $|01\rangle$ and $|10\rangle$, the coupling via $\hat{V}$ gives rise to an avoided crossing, with energy correction:
\begin{eqnarray}
\Delta E_1 = \frac{|\langle 10| \hat{V} |01\rangle|^2}{E_{01} - E_{10}} = \frac{g^2}{\Delta - \frac{3\epsilon^2}{4\alpha_2 - 2\delta_d}}.
\end{eqnarray}

For state $|11\rangle$, the leading contributions come from virtual transitions to $|E_2E_0\rangle$, $|E_0E_2\rangle$, and $|E_0E_3\rangle$:
\begin{eqnarray}
\Delta E_2 &&= \frac{|\langle 11| \hat{V} |E_2E_0\rangle|^2}{E_{11} - E_{20}} \nonumber\\
&&~~~~ + \frac{|\langle 11| \hat{V} |E_0E_2\rangle|^2}{E_{11} - E_{02}} + \frac{|\langle 11| \hat{V} |E_0E_3\rangle|^2}{E_{11} - E_{03}} \nonumber\\
&&= \frac{2g^2}{\Delta + \alpha_1 - \frac{3\epsilon^2}{4\alpha_2 - 2\delta_d}} \nonumber\\
&&~~~~+ \frac{g^2(1 - \delta_d/2r)}{\frac{\delta_d}{2} + r + \alpha_2 - \Delta - \frac{3\epsilon^2}{4\alpha_2 - 2\delta_d}} \nonumber\\
&&~~~~+ \frac{g^2(1 + \delta_d/2r)}{\frac{\delta_d}{2} - r + \alpha_2 - \Delta - \frac{3\epsilon^2}{4\alpha_2 - 2\delta_d}}.
\end{eqnarray}

Under the conditions $\Delta - \alpha_2 > 0$ and $\delta_d > \Delta - \alpha_2$, and keeping terms to order $\epsilon^2$, this simplifies to:
\begin{eqnarray}
\Delta E_2 \approx \frac{2g^2}{\Delta + \alpha_1} - \frac{2g^2}{(\Delta - \alpha_2) + \frac{3\epsilon^2}{\delta_d - \Delta + \alpha_2}}.
\end{eqnarray}

The effective ZZ coupling strength is obtained from the dressed energy levels:
\begin{eqnarray}
J_{zz} = \tilde{E}_{11} - \tilde{E}_{10} - \tilde{E}_{01} + \tilde{E}_{00} = \Delta E_2,
\end{eqnarray}
where $\tilde{E}_{ij}$ are the perturbed energies.

The expression for $J_{zz}$ reveals a remarkable drive dependence. For $\epsilon = 0$, we recover the static ZZ coupling:
\begin{eqnarray}
J_{zz}^{(0)} = \frac{2g^2}{\Delta + \alpha_1} - \frac{2g^2}{\Delta - \alpha_2} < 0.
\end{eqnarray}

As $\epsilon$ increases, the second term becomes modified, allowing $J_{zz}$ to be tuned. Including an additional correction from the $|13\rangle$ state (which becomes negligible due to the near-resonance between $|E_2\rangle$ and $|E_3\rangle$), we obtain the refined expression:
\begin{eqnarray}
J_{zz} = \frac{2g^2}{\Delta + \alpha_1} - \frac{2g^2}{(\Delta - \alpha_2) + \frac{3\epsilon^2}{\delta_d - \Delta + \alpha_2}} + \frac{3\epsilon^2}{4\alpha_2 - 2\delta_d}.
\label{eq:finalzz}
\end{eqnarray}

This function is monotonic in $\epsilon^2$ under our assumptions. The drive strength required to nullify the ZZ coupling is approximately:
\begin{eqnarray}
\epsilon_0 \approx \sqrt{\frac{\frac{1}{\Delta - \alpha_2} - \frac{1}{\Delta + \alpha_1}}{3\left[\frac{1}{2g^2(4\alpha_2 - 2\delta_d)} + \frac{1}{(\Delta - \alpha_2)^2(\delta_d - \Delta + \alpha_2)}\right]}}.
\end{eqnarray}

For $\epsilon > \epsilon_0$, $J_{zz}$ becomes positive, enabling complete control over the sign and magnitude of the ZZ interaction.


\section{APPENDIX E: ZZ Coupling Control in a Qubit-Coupler-Qubit Architecture}

The proposed parametric driving scheme can also be applied to a qubit-coupler-qubit architecture, where the coupler is driven to control the effective ZZ coupling between the two qubits. Here, we analyze the control mechanism and derive the parametric conditions for achieving a tunable ZZ interaction.

Consider a system where two qubits (Q1 and Q2) are coupled via a tunable coupler (C). The computational subspace is spanned by states $\ket{n_1, n_c, n_2}$, where $n_1$, $n_c$, and $n_2$ denote the excitation numbers of Q1, C, and Q2, respectively. Without parametric driving, the state $\ket{020}$ hybridizes with $\ket{110}$ and $\ket{011}$ due to static couplings, forming a dressed state $\widetilde{\ket{020}}$. This dressed state couples to $\ket{101}$ via an off-diagonal matrix element $\widetilde{\bra{020}}\hat{H}\ket{101} \neq 0$. As the coupler frequency $\omega_c$ varies, the energy of $\widetilde{\ket{020}}$ crosses that of $\ket{101}$, leading to a sign change in the effective ZZ coupling $J_{zz}$:
\begin{eqnarray}
\omega_c &&> \frac{\omega_1 + \omega_2 + \alpha_c}{2} \quad \Rightarrow \quad J_{zz} < 0, \nonumber\\
\omega_c &&< \frac{\omega_1 + \omega_2 + \alpha_c}{2} \quad \Rightarrow \quad J_{zz} > 0,
\end{eqnarray}
where $\omega_{1,2}$ are the qubit frequencies and $\alpha_c$ is the coupler anharmonicity.

Introducing a parametric drive on the coupler, $\hat{H}_d = \epsilon \cos(\omega_d t)(\hat{a}_c^{\dagger 2} + \hat{a}_c^2)$, resonantly couples $\ket{020}$ and $\ket{040}$, generating two dressed states $\widetilde{\ket{020}}$ and $\widetilde{\ket{040}}$ that interact with $\ket{101}$ via virtual transitions. Depending on the relative positions of $\widetilde{\ket{020}}$, $\widetilde{\ket{040}}$, and $\ket{101}$, the ZZ coupling exhibits distinct tuning behaviors as a function of drive strength $\epsilon$.

Let $\Delta_{101,020} = 2\omega_c - \omega_1 - \omega_2 - \alpha_c$ denote the energy detuning between $\ket{101}$ and the bare $\ket{020}$ state, and $\Delta_{101,040} = 4\omega_c - \omega_1 - \omega_2 - 6\alpha_c - \omega_d$ the detuning between $\ket{101}$ and the bare $\ket{040}$ state. The effective ZZ coupling $J_{zz}$ as a function of $\epsilon$ is determined by the competition between virtual transitions via $\widetilde{\ket{020}}$ and $\widetilde{\ket{040}}$. We identify four parametric regimes:

\noindent\textbf{Case 1:} $\omega_c > \frac{\omega_1 + \omega_2 + \alpha_c}{2}$ (static $J_{zz} < 0$).

\begin{enumerate}
\item If $\omega_d > 4\omega_c - \omega_1 - \omega_2 - 6\alpha_c$, then $\widetilde{\ket{020}}$ and $\widetilde{\ket{040}}$ lie on opposite sides of $\ket{101}$. Increasing $\epsilon$ monotonically drives $J_{zz}$ toward zero.

\item If $\omega_d < 4\omega_c - \omega_1 - \omega_2 - 6\alpha_c$, then both dressed states lie above $\ket{101}$. For weak driving, $J_{zz}$ becomes more negative; beyond a critical drive strength
\begin{eqnarray}
&&\epsilon_c = \\
&&\sqrt{\frac{(2\omega_c - \alpha_c - \omega_1 - \omega_2)(4\omega_c - 6\alpha_c - \omega_1 - \omega_2 - \omega_d)}{3}},\nonumber
\end{eqnarray}
$J_{zz}$ changes sign and continues to decrease.
\end{enumerate}

\noindent\textbf{Case 2:} $\omega_c < \frac{\omega_1 + \omega_2 + \alpha_c}{2}$ (static $J_{zz} > 0$).

\begin{enumerate}
\item If $\omega_d < 4\omega_c - \omega_1 - \omega_2 - 6\alpha_c$, then $\widetilde{\ket{020}}$ and $\widetilde{\ket{040}}$ lie on opposite sides of $\ket{101}$. Increasing $\epsilon$ monotonically drives $J_{zz}$ toward zero.

\item If $\omega_d > 4\omega_c - \omega_1 - \omega_2 - 6\alpha_c$, then both dressed states lie above $\ket{101}$. For weak driving, $J_{zz}$ increases further; beyond the critical drive strength $\epsilon_c$ given by Eq.~(3), $J_{zz}$ changes sign and continues to increase.
\end{enumerate}

These analytical predictions are derived from the conclusion of APPENDIX.~D. The critical drive strength $\epsilon_c$ marks the point where $\widetilde{\ket{040}}$ crosses $\ket{101}$, leading to a sign reversal of the ZZ coupling. This comprehensive parameter mapping enables precise control of $J_{zz}$, including complete suppression (zero coupling) and sign reversal, by appropriately choosing $\omega_c$, $\omega_d$, and $\epsilon$.
\end{document}